\documentclass[12pt,reqno,notitlepage]{amsart}

\usepackage[letterpaper,margin=1.25in]{geometry}

\usepackage{anyfontsize}
\usepackage[T1]{fontenc}

\usepackage{amsmath,amssymb,amsthm}

\usepackage{graphicx}
\usepackage[svgnames]{xcolor}
\usepackage{soul}
\usepackage{pgf}
\usepackage{tikz}
\usepackage{mathrsfs}
\usepackage{stmaryrd}
\SetSymbolFont{stmry}{bold}{U}{stmry}{m}{n}
\usepackage{layout}
\usepackage{mathtools}
\usepackage{paralist}
\usepackage{multirow}
\usepackage{comment}
\usepackage{rotating}
\usepackage{chngcntr}
\usepackage{apptools}

\usepackage{mathpazo}

\usepackage{xspace}
\usepackage{bm}
\usepackage{setspace}
\usepackage{accents}
\usepackage{enumitem}
\usepackage{footnote}
\usepackage{tabularx}
\usepackage{threeparttable}

\makesavenoteenv{tabular}
\makesavenoteenv{table}

\usepackage[colorlinks=true, linkcolor=Maroon, urlcolor=Maroon, citecolor=Maroon]{hyperref}

\usepackage[authoryear]{natbib}

\theoremstyle{plain}
\newtheorem{assumption}{Assumption}

\newtheorem{theorem}{Theorem}

\newtheorem{lemma}{Lemma}

\theoremstyle{remark}

\usepackage{todonotes}

\renewcommand{\Pr}{\mathbb{P}}

\newcommand{\Exp}{\mathbb{E}}

\usepackage{subfig}
\newcommand{\one}{\mathbb{1}}

\newcommand{\sC}{\mathcal{C}}

\usepackage{centernot}

\usepackage{cleveref}
\crefname{figure}{figure}{figures}
\creflabelformat{figure}{#2#1#3}

\crefname{equation}{equation}{equations}
\crefrangeformat{equation}{(#3#1#4) to~(#5#2#6)}
\creflabelformat{equation}{(#2#1#3)}

\crefname{lemma}{lemma}{lemmas}
\creflabelformat{lemma}{#2#1#3}
\crefname{proposition}{proposition}{propositions}
\creflabelformat{proposition}{#2#1#3}

\crefname{corollary}{corollary}{corollaries}
\creflabelformat{corollary}{#2#1#3}

\crefname{condition}{condition}{conditions}
\creflabelformat{condition}{#2#1#3}
\crefname{assumption}{assumption}{assumptions}
\creflabelformat{assumption}{#2#1#3}
\crefname{remark}{remark}{remarks}
\creflabelformat{remark}{#2#1#3}
\crefname{appendix}{appendix}{appendices}
\creflabelformat{appendix}{#2#1#3}

\allowdisplaybreaks
\begin{document}

\renewcommand{\thefootnote}{\fnsymbol{footnote}}

\begin{center}
\Large\textsc{Persuasion Effects in Regression Discontinuity Designs}
\vspace{1ex}
\end{center}

\vspace{.3cm}
\begin{center}
    \begin{tabular}{ccc}
        \large\textsc{Sung Jae Jun}\footnote{Department of Economics, suj14@psu.edu.} 
        &
        &
        \large\textsc{Sokbae Lee}\footnote{Department of Economics, sl3841@columbia.edu. }
        \\
        Pennsylvania State University
        &
        &
        Columbia University
    \end{tabular}
\end{center}

\begin{center}
\vspace{0.3cm} 
    September 30, 2025 
\vspace{0.3cm}
\end{center}

\noindent
\textbf{Abstract.}
We develop a framework for identifying and estimating persuasion effects in regression discontinuity (RD) designs. The \emph{RD persuasion rate} measures the probability that individuals at the threshold would take the action if exposed to a persuasive message, given that they would not take the action without exposure. We present identification results for both sharp and fuzzy RD designs, derive sharp bounds under various data scenarios, and extend the analysis to local compliers. Estimation and inference rely on local polynomial regression, enabling straightforward implementation with standard RD tools. Applications to public health and media illustrate its empirical relevance.





\raggedbottom
\clearpage


\renewcommand{\thefootnote}{\arabic{footnote}}
\setcounter{footnote}{0}


\section{Introduction}

Persuasive messaging plays a central role in shaping individual behavior, influencing actions across a wide range of domains, from health decisions to electoral choices. A growing empirical literature seeks to quantify the causal impact of such interventions. A natural metric is the \emph{persuasion rate}, defined as 
the probability that individuals would take the action if exposed to a persuasive message, given that they would not take the action without exposure.
While substantial progress has been made in identifying persuasion effects in randomized and observational settings since the foundational work of \citet{dellavigna2007fox}, formal results for regression discontinuity (RD) designs remain underdeveloped. This paper is to study the identification and estimation of persuasion rates in RD settings, where treatment assignment is determined---either deterministically or probabilistically---by whether a running variable crosses a threshold. 

We introduce and analyze the \emph{RD persuasion rate}, defined as the persuasion rate \emph{at the cutoff}. This parameter is analogous to the persuasion rate studied in \citet{jun2023identifying} but adapted to RD settings. We examine identification under both sharp and fuzzy designs. For the fuzzy design, we consider varying levels of observability for treatment status and assignment mechanisms, reflecting common data structures encountered in practice.  

In the sharp design, where treatment is a deterministic function of the running variable, we show that the RD persuasion rate is point identified under standard continuity and monotonicity assumptions. In the fuzzy design, where treatment is assigned probabilistically, we derive sharp bounds under three data scenarios: (i) the outcome, treatment status, and running variable are jointly observed; (ii) treatment status is unobserved, while the exposure rate is known; (iii) treatment status is unobserved, and no information for the exposure rate is available. The second and third scenarios are motivated by the fact that voting decisions are often observed only at the aggregate level, in which case the full joint distribution of the triplet is usually unavailable. Our results characterize how the identified set tightens with additional information on treatment status or the exposure rate function. 

In the fuzzy design, we further extend our analysis to a key subpopulation known as local compliers, i.e., units whose treatment status coincides with whether the running variable crosses the threshold within its local neighborhood. We provide identification results for the persuasion rate among local compliers under each of the three data scenarios. Under the most favorable scenario, the persuasion rate for local compliers is point identified and admits a practical interpretation as a Wald estimand. When the treatment status is not observed, the persuasion rate for local compliers shares the same sharp lower bound as the RD persuasion rate for the full population. 

Beyond identification, we develop estimation and inference procedures for both the sharp and fuzzy designs. In each case, the RD persuasion rate and its bounds can be estimated using local polynomial regression applied to standard or transformed outcome variables. We show how the delta method yields valid standard errors and explain how to construct confidence intervals that account for partial identification, building on the framework of \citet{Stoye:07}. Our procedures rely on widely used RD tools, such as the \texttt{rdrobust} package, and they are readily implementable in empirical applications.

We illustrate the empirical relevance of our framework with two applications. The first, based on \citet{brehm2025vaccines}, uses a sharp RD design to study the backlash effect of a landmark U.S.\ Supreme Court ruling on anti-vaccine newspaper discourse. The second, drawn from \citet{barone2015telecracy}, examines political behavior in Italy under a fuzzy RD design where media access varied discontinuously across regions. These examples demonstrate how our framework can be used to quantify persuasion effects in practice.

Although RD designs are widely used in applied research, their implications for persuasion analysis have not been systematically explored. The RD literature is extensive. Important early contributions include the formal identification results of \citet{hahn2001identification} and \citet{Lee2008JoE}'s influential application to estimating incumbency advantages in elections. Comprehensive overviews are available in the review articles by \citet{Imbens2008JoE}, \citet{Lee2010JEL}, and \citet{CattaneoTitiunik2022} as well as in the monographs by \citet{cattaneo2020foundations,cattaneo2024extensions}, which cover theoretical foundations, implementation strategies, and recent methodological developments.  

The persuasion rate, originally introduced by \citet{dellavigna2007fox} to assess the effectiveness of persuasive interventions, has become a central object of study across fields; see, for example, Table~1 in \citet{dellavigna2010persuasion} and Figure~7 in \citet{Bursztyn:Yang:22}. Building on this concept, \citet{jun2023identifying} formalize the persuasion rate as a causal parameter capturing behavioral responses to persuasive messages and provide identification results under exogenous treatment or valid instruments. Extending this work, \citet{jun2024aprt} introduce the forward and backward versions of the average persuasion rate on the treated, and they develop a difference-in-differences framework for their identification and estimation. Relatedly, \citet{jun2025tv} revisit the large-scale study of televised debates by \citet{TVdebate:23}, partially identifying persuasion parameters under monotonicity assumptions, without relying on experimental or quasi-experimental variation. Other recent contributions include the classification of persuasion types among compliers \citep{yu2023binary}, identification under sample selection \citep{possebom2022probability}, extensions to continuous outcomes \citep{kaji2023assessing}, and covariate-assisted bounds without monotonicity assumptions \citep{ji2023model}. To the best of our knowledge, however, there is no existing work that formally analyzes the persuasion rates in RD designs.

Overall, our results offer a unified framework for persuasion analysis in RD designs. The identified quantities and estimators are transparent, flexible, and easy to implement. This paper contributes to the methodological foundations of recent research on the persuasion rates and provides applied researchers with practical tools for measuring the impact of persuasive efforts at the cutoff.

The remainder of the paper is organized as follows. In \Cref{sec:iden}, we formalize the RD persuasion rate and establish identification results under both sharp and fuzzy designs across three data scenarios. \Cref{sec:estimation} presents estimation and inference procedures based on local polynomial regression, including methods for constructing valid standard errors and confidence intervals. In \Cref{sec:examples}, we apply our methodology to two empirical settings: a U.S.\ Supreme Court ruling on vaccine mandates and a natural experiment involving media exposure in Italian politics. These examples illustrate the practical applicability of our methods. 
\Cref{sec:concl} gives concluding remarks, and \cref{sec:proofs} contains proofs omitted from the main text.

\section{Identification}\label{sec:iden}

Let $Y(1)$ and $Y(0)$ denote binary potential outcomes, and let $D$ represent a binary treatment. In the context of persuasion, $D$ typically corresponds to an informational treatment, that is, exposure to a persuasive message. The observed outcome is given by $Y = Y(1)D + Y(0)(1 - D)$. We may also observe a vector $X$ of covariates; however, we omit it from our discussion for brevity, as it does not play an important role in the analysis.

We consider a RD design framework in which a continuously distributed running variable $W$ is observed. Following common practice in the literature, we normalize the known threshold to zero. Treatment assignment may be deterministic, as in $D = \one(W \geq 0)$ with probability one, where $\one( \cdot )$ denotes the usual indicator function, or it may be probabilistic in that 
\[
\Pr(D = 1 \mid W) = e_p(W)\one(W \geq 0) + e_n(W)\one(W < 0),
\]
where $e_n(0) < e_p(0)$; the subscripts $n$ and $p$ are chosen to indicate the negative and positive sides, respectively. The former case corresponds to a \emph{sharp} design, while the latter defines a \emph{fuzzy} design.

Our goal is to understand the identification of
\begin{equation}\label{eq:lpr} 
\theta := \Pr\{ Y(1) = 1 \mid Y(0) = 0, W = 0 \},
\end{equation}
provided that this conditional probability is well-defined. We refer to $\theta$ as the \emph{persuasion rate at the cutoff}, or simply the \emph{RD persuasion rate}.

The interpretation of $\theta$ parallels that of the causal persuasion rate studied by \citet{jun2023identifying}. Conditioning on individuals at the threshold $W = 0$, it captures the share of those who would take the action of interest when exposed to a persuasive message among those who would not otherwise. In this way, we avoid ``preaching to the converted'' by focusing on the target subpopulation the persuasive message aims to influence. The persuasive effect is thus measured by the extent to which individuals change their action in response to the message.

Throughout the paper, we consider the following assumptions.

\begin{assumption}[Monotone Treatment Response] \label{ass:mtr}
$Y(1) \geq Y(0)$ with probability one.
\end{assumption}

\begin{assumption}[Well-Defined RD Persuasion Rate] \label{ass:overlap}
$\Pr\{ Y(0) = 0 \mid W = 0 \} > 0$.
\end{assumption}

\Cref{ass:overlap} ensures that the parameter $\theta$ in \eqref{eq:lpr} is well-defined. Monotonicity assumptions such as \cref{ass:mtr} are widely used in the literature; see, for example, \citet{manski1997monotone}, \citet{manski2000mtr}, \citet{pearl1999probabilities}, and \citet{jun2023identifying,Jun2024JBES,jun2024aprt}, among others. In our context, this assumption reflects the directional (or biased) nature of persuasive messages, a common feature in persuasion settings. It rules out backlash effects, where exposure to the message causes a shift in the opposite direction due to distrust, annoyance, or perceived coercion. For instance, overly forceful or manipulative messages may provoke resistance rather than compliance. Such counterproductive responses are excluded under \cref{ass:mtr}.

The RD persuasion $\theta$ depends on the joint distribution of the potential outcomes that are never observed simultaneously. Therefore, we also consider the following quantity that depends only on the marginal distributions of the potential outcomes: 
\[
\theta_L := \frac{\Pr\{ Y(1) = 1 \mid W = 0 \} - \Pr\{ Y(0) = 1 \mid W = 0 \}}{1 - \Pr\{ Y(0) = 1 \mid W = 0 \}},
\] 
which is well-defined whenever \cref{ass:overlap} holds. 
The following lemma, whose unconditional version appears as Lemma~1 in \citet{jun2023identifying}, characterizes the extent to which $\theta$ can be expressed or bounded based on the marginals of the potential outcomes, depending on whether \cref{ass:mtr} is imposed.

\begin{lemma}\label{lem:ate vs pr}
Suppose that \cref{ass:overlap} hold. If \cref{ass:mtr} holds, then $\theta = \theta_L$. Otherwise, $\theta$ satisfies the sharp bounds 
\begin{equation}\label{lemma1-result0}
\max\left\{ 0,\ \theta_L \right\}
\;\leq\; \theta \;\leq\;
\min\left\{ 
\frac{\Pr\{ Y(1) = 1 \mid W = 0 \}}{1 - \Pr\{ Y(0) = 1 \mid W = 0 \}},\ 
1
\right\},
\end{equation}
where the sharpness means that any value within this interval is compatible with the marginal distributions of the potential outcomes. 
\end{lemma}

The inequalities in \eqref{lemma1-result0} follow from the Fr\'echet--Hoeffding bounds on joint distributions with fixed marginals. If \cref{ass:mtr} is imposed, then there is no difference between $\theta$ and $\theta_L$, because $\Pr\{ Y(0) = 1\mid W=0\} = \Pr\{ Y(1) =1, Y(0) = 1 \mid W=0\}$. 

 Since $Y(1)$ and $Y(0)$ are never observed simultaneously, the inequalities in \eqref{lemma1-result0} represent the tightest bounds we can obtain for studying $\theta$ without additional assumptions. Further, if monotonicity is imposed---a plausible condition in the context of persuasive messages---these bounds can be substantially sharpened to a single point, $\theta_L$. Therefore, although \cref{lem:ate vs pr} is not yet an identification result, it highlights the central role of $\theta_L$ in analyzing $\theta$: under monotonicity, $\theta_L = \theta$, and more generally, $\theta_L$ serves as a robust lower bound on $\theta$.

\subsection{The Sharp Design}

We begin with the sharp design case.

\begin{assumption}[Sharp Design]\label{ass:sharp}
We have $D = \one(W \geq 0)$ with probability one.
\end{assumption}

That is, treatment assignment is a deterministic function of the running variable $W$. Define
\begin{equation} \label{eq:srdd}
\theta_{\mathrm{RD}} :=  \frac{\Pr( Y = 1 \mid W = 0^+ ) - \Pr( Y = 1 \mid W = 0^- )}{1 - \Pr( Y = 1 \mid W = 0^- )},
\end{equation}
where
\begin{align*}
    \Pr(Y = 1 \mid W = 0^+) &:= \lim_{h \downarrow 0} \Pr(Y = 1 \mid W = h), \\
    \Pr(Y = 1 \mid W = 0^-) &:= \lim_{h \uparrow 0} \Pr(Y = 1 \mid W = h).
\end{align*}
Throughout the paper, we use $0^+$ and $0^-$ to denote the right- and left-hand limits, respectively, and we assume that these limits exist and are well-defined.

As in standard RD analysis, a simple continuity condition allows us to connect $\theta_{\mathrm{RD}}$ with  $\theta_L$.

\begin{assumption}[Continuity of Counterfactual Outcome Probabilities]\label{ass:continuity}
For each $d \in \{0,1\}$, the function $w \mapsto \Pr\{ Y(d) = 1 \mid W = w \}$ is continuous at $w = 0$.
\end{assumption}

\begin{theorem} \label{thm:sharp}
Suppose that \cref{ass:mtr,ass:overlap,ass:sharp,ass:continuity} hold. Then, $\theta$ is point identified by $\theta_{\mathrm{RD}}$. Moreover, without \cref{ass:mtr}, the bounds in \eqref{lemma1-result0} are still identified, and $\theta_L = \theta_{\mathrm{RD}}$ holds.
\end{theorem}

\Cref{thm:sharp} shows that in the sharp design, the RD persuasion rate $\theta$ is identified from observed data under standard smoothness and monotonicity assumptions. Specifically, the discontinuity in the outcome probability at the threshold captures the RD treatment effect, which is then rescaled to reflect the subpopulation of individuals who would not act in the absence of persuasion. Without \cref{ass:mtr}, the same formula identifies a lower bound on $\theta$.

\subsection{The Fuzzy Design}

We now turn to the case where treatment assignment is probabilistic rather than deterministic. Specifically, the exposure probability \( e(W) := \Pr(D = 1 \mid W) \) is not a simple step function, but it still has a discontinuity point at $W=0$, with units just above the cutoff more likely to receive treatment than those just below. To formalize this setting, we follow the approach of \citet{hahn2001identification} and 
adopt the threshold-crossing framework of \citet{vytlacil2002independence}.
Let $D(w)$ denote the potential treatment status when the running variable $W$ takes the (potential) value $w$.

\begin{assumption}[Fuzzy Design] \label{ass:threshold} 
(i) $D \equiv D(W)$, where $D(w) = \one\{ V \leq e(w) \}$ and $V$ is uniformly distributed on $[0,1]$. The function $e(w)$ takes the form
\[
e(w) = e_p(w)\one(w \geq 0) + e_n(w)\one(w < 0),
\]
where $e_p$ and $e_n$ are continuous at $0$ and satisfy $e_p(h) > e_n(-h)$ for all sufficiently small $h > 0$.
(ii) $\bigl(Y(1), Y(0), V\bigr)$ is jontly independent of $W$ in a neighborhood of $W = 0$. 
\end{assumption}

\Cref{ass:threshold}(ii) is a local independence condition, while \cref{ass:threshold}(i) ensures a discontinuity in the exposure rate at the cutoff and rules out defiers in a neighborhood of $W = 0$.\footnote{The function $e(w) = \Pr(D = 1 \mid W = w)$ is referred to as the exposure rate to emphasize the persuasion context; in the broader literature, it is typically called the propensity score.}
In this setting, compliers at the threshold correspond to units with $V \in (e(0^-), e(0^+)]$, and we will refer to them as \emph{local compliers}.

Define
\begin{align*}  
    \theta_{\mathrm{RD},U} 
    &:= 
    \frac{\Pr(Y = 1, D = 1 \mid W = 0^+) + 1 - e(0^+) - \Pr(Y = 1, D = 0 \mid W = 0^-)}{1 - \Pr(Y = 1, D = 0 \mid W = 0^-)}, \\
    \theta_{\mathrm{RD},U,e} 
    &:= 
    \frac{\min\{1,\ \Pr(Y = 1 \mid W = 0^+) + 1 - e(0^+)\} - \max\{0,\ \Pr(Y = 1 \mid W = 0^-) - e(0^-)\}}{1 - \max\{0,\ \Pr(Y = 1 \mid W = 0^-) - e(0^-)\}},
\end{align*} 
and we have the following results. 

\begin{theorem}\label{thm:fuzzy}
Suppose that \cref{ass:mtr,ass:overlap,ass:continuity,ass:threshold} hold.
\begin{enumerate}
\item If $(Y, D, W)$ is jointly observed, then the sharp identified bounds on $\theta = \theta_L$ are given by $[\theta_{\mathrm{RD}},\ \theta_{\mathrm{RD},U}]$. 

\item If only $(Y, W)$ is observed, but $e(W)$ is known from an external source, then the sharp identified bounds on $\theta = \theta_L$ are given by $[\theta_{\mathrm{RD}},\ \theta_{\mathrm{RD},U,e}]$. 

\item If only $(Y, W)$ is observed and no information about $e(W)$ is available, then the sharp identified bounds on $\theta = \theta_L$ are given by $[\theta_{\mathrm{RD}},\ 1]$.
\end{enumerate}
\end{theorem}

\Cref{thm:fuzzy} highlights how the identification of the persuasion rate $\theta$ depends on the availability of treatment-related information. When the full triplet $(Y, D, W)$ is observed, as in case (i), the sharp lower bound $\theta_{\mathrm{RD}}$ can be computed without using $D$, while the sharp upper bound $\theta_{\mathrm{RD},U}$ incorporates additional information from observable population quantities such as $\Pr(Y = 1, D = 1 \mid W = 0^+)$, $\Pr(Y = 1, D = 0 \mid W = 0^-)$ and the exposure rate $e(0^+)$. However, in many voting-related examples, voting decisions are rarely observed at the individual level, and therefore they are usually not observed jointly with individual exposures to a political message. If the researcher has access to aggregate level data on $(Y,W)$ and $(D,W)$ separately, then it corresponds to case (ii), in which case the lower bound remains unchanged, while the upper bound is conservatively adjusted to account for the missing treatment-status information. If the researcher has access only to the distribution of $(Y,W)$, then we are in case (iii), the most restricted setting.  In this case, the upper bound defaults to the logical maximum of $1$, while the lower bound $\theta_{\mathrm{RD}}$ still remains the same. Across all three scenarios, the lower bound remains robust, while the upper bound becomes tighter as more treatment-related information becomes available. Therefore, when recovering the joint distribution of the triplet $(Y,D,W)$ is costly due to limited data access, as in voting applications, there is no loss of information for the purpose of identifying the sharp lower bound in restricting attention to the joint distribution of $(Y,W)$ alone.

\subsection{The Fuzzy Design and Local Compliers}

In the fuzzy design, one may wish to focus on the group of compliers. In this case, a natural parameter of interest is the persuasion rate for compliers at $W = 0$, i.e., \emph{local compliers}. Specifically, define 
\begin{equation}
    \theta_c := \Pr\left\{ Y(1) = 1 \mid Y(0) = 0,\ e(0^-) < V \leq e(0^+) \right\},
\end{equation}
which is well-defined under the following assumption. Here, the subscript $c$ denotes the compliers.

\begin{assumption}[Well-Defined RD Persuasion Rate for Local Compliers]\label{ass:overlap-compliers}
\[
\Pr\left\{ Y(0) = 0 \mid e(0^-) < V \leq e(0^+) \right\} > 0.
\]
\end{assumption}

Since \( \theta_c \) depends on the joint distribution of the potential outcomes, it is not point identified from observed data in general.  As in the sharp design, this issue can be addressed using \cref{ass:mtr}. Let \( \sC_0 \) denote the group of compliers local to the threshold, characterized by \( e(0^-) < V \leq e(0^+) \). Define
\[
\theta_{cL} 
:= 
\frac{ \Pr\left\{ Y(1) = 1 \mid \sC_0 \right\} - \Pr\left\{ Y(0) = 1 \mid \sC_0 \right\} }{ 1 - \Pr\left\{ Y(0) = 1 \mid \sC_0 \right\}},
\]
which serves as a lower bound on \( \theta_c \) in the absence of further assumptions.

The relationship between $\theta_c$ and $\theta_{cL}$ is similar to that of $\theta$ and $\theta_L$. Specifically, the persuasion rate for compliers $\theta_c$ can be generally bounded by objects that depend only on the marginal distributions of the potential outcomes. Those bounds will take the same functional form as in \cref{lem:ate vs pr}, replacing the whole population at the cutoff with the local complier group $\sC_0$: i.e., $\theta_{cL}$ is a robust lower bound on $\theta_c$ in general. Of course, the bounds collapse to a single point, i.e., $\theta_c = \theta_{cL}$, if and only if \cref{ass:mtr} holds.  Therefore, $\theta_{cL}$ plays a central role in the analysis of $\theta_c$. 

We now formally state our identification results for $\theta_{cL}$. Define 
\begin{align*}  
    \theta_{cL}^* 
    &:= 
    \frac{ \Pr(Y = 1 \mid W = 0^+) - \Pr(Y = 1 \mid W = 0^-) }{ \Pr(Y = 0,\ D = 0 \mid W = 0^-) - \Pr(Y = 0,\ D = 0 \mid W = 0^+) }, \\
    \theta_{cL}^{**}
    &:= 
    \max\left\{ \theta_{\mathrm{RD}},\ \frac{ \Pr(Y = 1 \mid W = 0^+) - \Pr(Y = 1 \mid W = 0^-) }{ e(0^+) - e(0^-) } \right\}.
\end{align*}

\begin{theorem}\label{thm:fuzzy-compliers}
Suppose that \cref{ass:mtr,ass:continuity,ass:threshold,ass:overlap-compliers} hold.
\begin{enumerate}
\item If $(Y, D, W)$ is jointly observed, then $\theta_c = \theta_{cL}$ is point identified by $\theta_{cL}^*$.

\item If only $(Y, W)$ is observed and the exposure rate function $e$ is known from an external source, then the sharp identified bounds on $\theta_c = \theta_{cL}$ are given by $[\theta_{cL}^{**},\ 1]$.

\item If only $(Y, W)$ is observed and no information about $e$ is available, then the sharp identified bounds on $\theta_c = \theta_{cL}$ are given by $[\theta_{\mathrm{RD}},\ 1]$.
\end{enumerate}
\end{theorem}

In case (i), the persuasion rate \( \theta_c \) for local compliers is point identified by \( \theta_{cL}^* \). The numerator of \( \theta_{cL}^* \) captures the jump in the outcome probability at the threshold, while the denominator reflects the change in the share of untreated individuals who would not take the action of interest in the absence of persuasion. This expression directly leverages the joint observability of $(Y, D, W)$. In case (ii), where treatment assignment $D$ is not observed jointly with $(Y,W)$ but the exposure rate function $e$ is known from an external source, the lower bound $\theta_{cL}^{**}$ improves upon $\theta_{\mathrm{RD}}$. This expression combines the observed jump in outcome probability with externally provided information on the change in treatment probability, potentially yielding an improved lower bound. In case (iii), where only $(Y, W)$ is observed, and no information about the exposure rate is available, the lower bound reverts to $\theta_{\mathrm{RD}}$, the most conservative estimate based solely on the outcome discontinuity. In the absence of any knowledge about treatment assignment probabilities, restricting ourselves to the group of local compliers does not provide anything better than $\theta_{\mathrm{RD}}$ as a conservative measure. The upper bound remains at $1$, reflecting the logical maximum.


\subsection{Relation to Probabilities of Causation}

The persuasion rate is conceptually related to the notion of probabilities of causation;
see, for example, \citet{pearl1999probabilities,tian2000probabilities,Yamamoto:2012,Dawid2014fitting,Dawid2022effects,Ding:2024:arXiv:prob_necessity}. A key feature of this literature, particularly following the foundational work of \citet{pearl1999probabilities}, is the emphasis on defining causal probabilities by conditioning on observable quantities rather than counterfactual ones. This leads to expressions such as the \emph{probability of sufficiency}, $\mathrm{PS} := \mathbb{P}(Y(1) = 1 \mid Y = 0, D = 0)$, and the \emph{probability of necessity}, $\mathrm{NS}:=\mathbb{P}(Y(0) = 0 \mid Y = 1, D = 1)$. These formulations avoid conditioning on unobservable potential outcomes and instead rely on observed variables. To the best of our knowledge, there is no systematic study of these probabilities of causation within the RD framework. Developing such analogs remains an open direction for future research.

\section{Estimation and Inference}\label{sec:estimation}

Estimation and inference in RD designs have been extensively studied. Local polynomial estimation is the standard method for point estimation. For statistical inference, the most common approach is \emph{bias-corrected} inference, which is based on a bias-corrected RD estimator \citep[e.g.,][]{calonico2014robust}. An alternative approach, known as \emph{bias-aware} inference, constructs confidence intervals that explicitly account for the worst-case bias of nonparametric estimators \citep[e.g.,][]{ArmstrongKolesar2018,ArmstrongKolesar2020,ImbensWagerr2019}. In our context where a typical estimand is a ratio of nonparametric estimators, this amounts to constructing confidence sets akin to Anderson-Rubin-type intervals, in order to address issues such as the failure of the delta method under weak identification \citep[e.g.,][]{KolesarRothe2018,NoackRothe2024}. In this paper, we adopt the standard bias-corrected inference method, which is implemented in the widely used \texttt{rdrobust} package for Stata \citep{calonico2017rdrobust}, with R and Python versions available at \url{https://rdpackages.github.io/rdrobust/}. The development of alternative inference methods, including bias-aware procedures, remains an important direction for future research.\footnote{Inference for RD designs remains an active area of research. For instance, \citet{PLRD} introduce a new method, called partially linear regression discontinuity inference, to address limitations of existing bias-corrected and bias-aware approaches under a stronger assumption on treatment effects.}

\subsection{The Sharp Design}

In the sharp design case, both parametric and nonparametric approaches are possible. We advocate nonparametric methods, but we briefly discuss parametric options for completeness. 

The easiest (but restrictive) approach is probably the one using a linear-in-coefficients probability model and ordinary least squares (OLS). To illustrate the idea, suppose that for $d \in \{0,1\}$ and some pre-specified integer $J$, 
\begin{equation}\label{eq:linear parametric}
\Pr\{ Y(d) = 1\mid W = w \} = \sum_{j=0}^J \alpha_{jd} w^j. 
\end{equation}
Then,
\[
\theta_L = \frac{\alpha_{01} - \alpha_{00}}{1-\alpha_{00}}. 
\]
Since $D$ is deterministically assigned based solely on $W$ in the sharp design, the conditional expectation of the observed outcome can now be expressed as:
\begin{align*} 
\Exp(Y \mid D, W)
&=
D \Exp\{ Y(1) \mid W \} + (1-D) \Exp\{ Y(0) \mid W \} \\
&= 
\alpha_{00} + (\alpha_{01} - \alpha_{00}) D + \sum_{j=1}^J \left\{ \alpha_{j0} W^j + (\alpha_{j1} - \alpha_{j0}) D W^j \right\}.
\end{align*} 
Therefore, in a linear parametric setup like \eqref{eq:linear parametric}, we can simply run OLS of $Y$ on the intercept, $D$, polynomials of $W$, and their interactions with $D$, after which we can obtain 
\[
\theta_L = \frac{\text{coefficient on } D}{1 - \text{intercept}}. 
\]
We can then find the standard error by using the variance-covariance matrix of the OLS estimates and the delta method. 

While the linear-in-coefficients model offers convenience, it imposes restrictive functional form assumptions. Replacing \eqref{eq:linear parametric} with nonlinear parametric models would simply involve estimating two separate nonlinear regressions: one regression of $Y$ on $W$ for units with $D=0$, and the other for those with $D=1$. The standard error of the resulting estimate of $\theta_L$ can then be readily computed using the delta method as the two regressions are based on two independent samples in the i.i.d.\ setting.  

A more general approach is the one that avoids parametric assumptions altogether.  To this end, we recommend that we estimate $\theta_{\mathrm{RD}}$, defined in \eqref{eq:srdd}, by local polynomial regression. This is the approach we advocate, and we will elaborate below. To simplify notation, let
\[
    \mu_+ := \Pr(Y = 1 \mid W = 0^+) \quad \text{and} \quad
    \mu_- := \Pr(Y = 1 \mid W = 0^-),
\]
and refer to them as the \textit{right} and \textit{left} estimands, respectively. Then,
\begin{equation}\label{eq:thetaRD local}
\theta_{\mathrm{RD}} = \frac{\mu_+ - \mu_-}{1 - \mu_-},
\end{equation}
where the numerator $\mu_+ - \mu_-$ corresponds to the standard RD treatment effect under the sharp design.  

Estimating $\mu_+$ and $\mu_-$ by local polynomial regression requires that we select bandwidths that define the neighborhood around the threshold, which may be chosen symmetrically or asymmetrically for the left and right side. Let $\hat{\mu}_+$ and $\hat{\mu}_-$ denote the local polynomial regression estimates from the right and left sides, respectively, typically reported by RD packages (e.g., in Stata, \texttt{rdrobust} saves them as output variables).
We define the estimator of $\theta_{\mathrm{RD}}$ as
\begin{equation}\label{eq:thetahat RD}
\hat{\theta}_{\mathrm{RD}} := \frac{\hat{\mu}_+ - \hat{\mu}_-}{1 - \hat{\mu}_-}.
\end{equation}
To compute the standard error for $\hat{\theta}_{\mathrm{RD}}$, we apply the delta method, assuming that $\hat{\mu}_+$ and $\hat{\mu}_-$ are asymptotically independent. This assumption holds in i.i.d.\ settings where the two estimates are based on disjoint subsamples. Let $\widehat{\mathrm{se}}_+$ and $\widehat{\mathrm{se}}_-$ denote the standard errors of $\hat{\mu}_+$ and $\hat{\mu}_-$, respectively. In \texttt{rdrobust}, these correspond to the leading diagonal entries of the variance-covariance matrices \texttt{e(V\_cl\_r)} and \texttt{e(V\_cl\_l)} for the conventional estimator, or \texttt{e(V\_rb\_r)} and \texttt{e(V\_rb\_l)} for the robust estimator.
Then, the standard error of $\hat{\theta}_{\mathrm{RD}}$ is given by
\begin{align}\label{se-formula:delta}
\widehat{\mathrm{se}}(\hat{\theta}_{\mathrm{RD}}) = \left[ 
\left( \frac{1}{1 - \hat{\mu}_-} \right)^2 \widehat{\mathrm{se}}_+^2 + 
\left( \frac{\hat{\mu}_+ - 1}{(1 - \hat{\mu}_-)^2} \right)^2 \widehat{\mathrm{se}}_-^2 
\right]^{1/2}.
\end{align}
This formula applies to both the conventional and bias-corrected estimators produced by \texttt{rdrobust}. 

In sum, our estimation and inference procedures build directly on standard tools for the sharp RD design. While the optimal bandwidth and robust inference procedures developed for standard RD estimands are not guaranteed to be optimal for our new estimand $\theta_{\mathrm{RD}}$, they are likely to perform well in practice, provided that the estimated denominator $(1 - \hat{\mu}_-)$ is not too close to zero. 
A formal study of the optimality of these procedures for our estimand remains an important direction for future research.

\subsection{The Fuzzy Design}

We focus on the data setting in which $(Y, D, W)$ is jointly observed, as it is the most informative and allows for straightforward implementation of the sample analog estimation methods.  We focus on nonparametric approaches based on local polynomials. 
In the fuzzy design, $\theta = \theta_L$ is only partially identified. The sharp bounds can be estimated by using the sample analog principle, but inference requires caution.  

To be more specific, recall from \Cref{thm:fuzzy}~(i) that the sharp identifiable bounds on $\theta_L$ in the current setup are given by $[\theta_{\mathrm{RD}},\ \theta_{\mathrm{RD},U}]$, where $\theta_{\mathrm{RD}}$ is the same as in the case of the sharp design: see \cref{eq:thetaRD local}. Therefore, $\theta_{\mathrm{RD}}$ can be estimated as in \eqref{eq:thetahat RD}, and its standard error can be computed by \eqref{se-formula:delta}. The upper bound $\theta_{\mathrm{RD},U}$ can be expressed as 
\[
\theta_{\mathrm{RD},U} = \frac{\mu_{U,+} - \mu_{U,-}}{1 - \mu_{U,-}},
\]
where
\[
\mu_{U,+} := \mathbb{E}[YD + 1 - D \mid W = 0^+]
\quad \text{and} \quad
\mu_{U,-} := \mathbb{E}[Y(1 - D) \mid W = 0^-].
\]
These quantities can be estimated using standard RD procedures again: $\mu_{U,+}$ is obtained by treating $(YD + 1 - D)$ as the outcome and estimating the right-hand limit at the threshold, while $\mu_{U,-}$ is estimated analogously using $Y(1 - D)$ from the left. Once the standard errors of $\mu_{U,+}$ and $\mu_{U,-}$ are obtained, the standard error of $\hat\theta_{\mathrm{RD},U}$ can be computed using the delta method, as described in \eqref{se-formula:delta}.

Since $\theta_L$ lies in the interval $[\theta_{\mathrm{RD}},\ \theta_{\mathrm{RD},U}]$, inference must account for this partial identification. The method proposed by \citet{Stoye:07} offers a valid approach to constructing confidence intervals in this setting. Specifically, a $(1 - \alpha)$ confidence interval for $\theta$ is given by
\[
\left[
\hat\theta_{\mathrm{RD}} - c_\alpha\, \widehat{\mathrm{se}}(\hat\theta_{\mathrm{RD}}),\quad
\hat\theta_{\mathrm{RD},U} + c_\alpha\, \widehat{\mathrm{se}}(\hat\theta_{\mathrm{RD},U})
\right],
\]
where $\widehat{\mathrm{se}}(\hat\theta_{\mathrm{RD}})$ and $\widehat{\mathrm{se}}(\hat\theta_{\mathrm{RD},U})$ denote the estimated standard errors of the lower and upper bounds, respectively. The critical value $c_\alpha$ is determined by solving
\[
\Phi\left( c_\alpha + \frac{\hat\Delta}{\max\left\{ \widehat{\mathrm{se}}(\hat\theta_{\mathrm{RD}}),\ \widehat{\mathrm{se}}(\hat\theta_{\mathrm{RD},U}) \right\}} \right) - \Phi(-c_\alpha) = 1 - \alpha,
\]
where $\Phi$ is the cumulative distribution function of the standard normal distribution, and $\hat\Delta := \hat\theta_{\mathrm{RD},U} - \hat\theta_{\mathrm{RD}}$ denotes the estimated width of the identified interval.

We now turn to the estimation of $\theta_{cL}^*$, which is point identified as
\[
\theta_{cL}^* 
= 
\frac{ \Pr(Y = 1 \mid W = 0^+) - \Pr(Y = 1 \mid W = 0^-) }{ \Pr(Y = 0,\ D = 0 \mid W = 0^-) - \Pr(Y = 0,\ D = 0 \mid W = 0^+) },
\]
as shown in \cref{thm:fuzzy-compliers}~(i). This expression can be interpreted as a Wald estimand by treating the variable $Y + D - YD$ as a pseudo-treatment indicator, which equals 0 if both \( Y = 0 \) and \( D = 0 \), and 1 otherwise.
Indeed, the denominator can be rewritten as
\begin{align*}
&\Pr(Y = 0,\ D = 0 \mid W = 0^-) - \Pr(Y = 0,\ D = 0 \mid W = 0^+) \\
&= \mathbb{E}[Y + D - YD \mid W = 0^+] - \mathbb{E}[Y + D - YD \mid W = 0^-].
\end{align*}
In other words, estimation of $\theta_{cL}^*$ can proceed using a standard fuzzy RD estimator applied to the outcome variable $Y$ and the pseudo-treatment variable $(Y + D - YD)$. This allows for direct implementation using off-the-shelf software, simply by redefining the treatment indicator accordingly.

We now briefly comment on the other two data scenarios. First, if only the joint distribution of $(Y, W)$ is observed, then the sharp identifiable bounds on both $\theta_L$ and $\theta_{cL}$ become $[\theta_{\mathrm{RD}},\ 1]$. In this case, inference can proceed using a one-sided confidence interval for $\theta_{\mathrm{RD}}$, treating it as a lower bound. Second, consider the intermediate case in which only $(Y, W)$ is observed but the function $e(W)$ is estimated from an external source. While plug-in estimation of the bounds remains straightforward, valid inference becomes more challenging due to the use of two separate samples and the non-smooth nature of the bounds. \citet[online Appendices F and G]{jun2023identifying} develop inference procedures for such settings in the context of instrumental variable estimation. As the arguments are extendable but tedious to reproduce, we refer interested readers to those appendices for further details.

\subsection{Decision Flow for Estimation and Inference}

\begin{figure}[!htbp]
\centering
\includegraphics[width=1.00\textwidth]{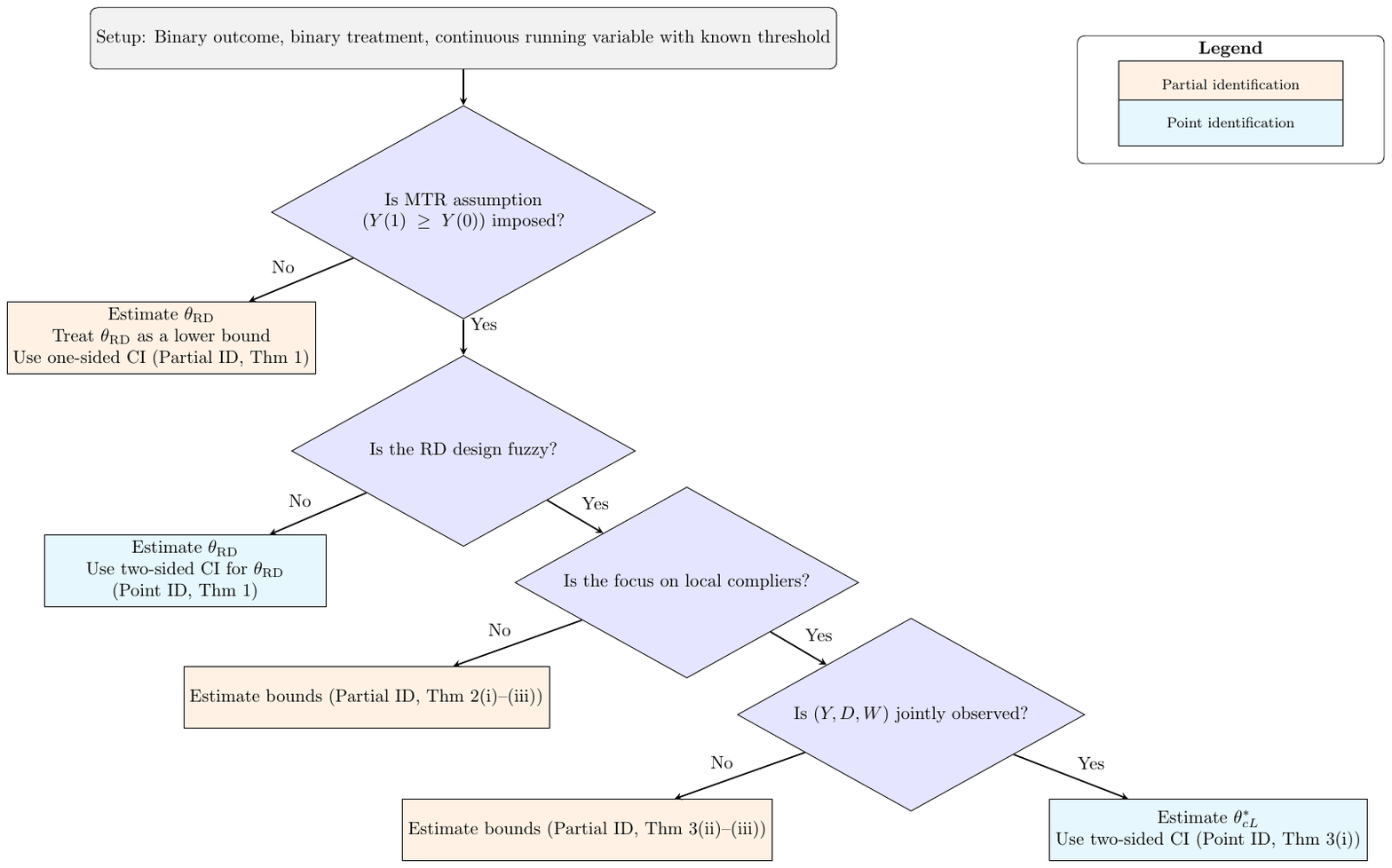}
\caption{Flow Chart for Estimation and Inference}
\label{fig:rd-flowchart}
\end{figure}

The estimation and inference procedures discussed above are summarized in the flow chart in \cref{fig:rd-flowchart}, which outlines a decision-making process based on key design features and identifying assumptions. Below we discuss each of the decision nodes in detail.   The setup follows a standard RD design with the additional assumption that the outcome variable is binary. The causal parameter of interest is $\theta$, the persuasion rate at the threshold.

The first decision is whether to impose \cref{ass:mtr}, i.e., the MTR assumption.  If MTR is not imposed, then the estimand $\theta_{\mathrm{RD}}$ is interpreted as a lower bound on $\theta$: specifically, $\theta_{\mathrm{RD}} = \theta_L\leq \theta$ in the sharp design and $\theta_{\mathrm{RD}}\leq \theta_L \leq \theta$ in the fuzzy design. Therefore, inference should proceed with a one-sided confidence interval, $[\hat\theta_{\mathrm{RD}} - z_{1-\alpha}\, \widehat{\mathrm{se}}(\hat\theta_{\mathrm{RD}}),\; 1]$, where $z_{1-\alpha}$ denotes the $(1 - \alpha)$ quantile of the standard normal distribution.

If MTR is imposed, then the distinction between sharp and fuzzy RD designs becomes central. In the sharp design, $\theta$ is point identified by $\theta_{\mathrm{RD}}$, which can be estimated using standard methods. In this case, inference proceeds using a two-sided confidence interval, i.e., $[\hat\theta_{\mathrm{RD}} - z_{1-\alpha/2}\, \widehat{\mathrm{se}}(\hat\theta_{\mathrm{RD}}),\; \hat\theta_{\mathrm{RD}} + z_{1-\alpha/2}\, \widehat{\mathrm{se}}(\hat\theta_{\mathrm{RD}})]$. In the fuzzy design, we should decide whether we want to keep $\theta$ as a parameter of interest, or we want to change the target to focus on the subpopulation of local compliers. In the former approach, we have partial identification of $\theta$, where the exact form of the sharp identified bounds on $\theta$ depends on how $D$ is observed.  If we are willing to focus on local compliers, then we can consistently estimate $\theta_{cL}^*$ if we have data on the triplet $(Y,D,W)$; otherwise, $\theta_{cL}^*$ is only partially identified.

Figure 1 summarizes this reasoning and provides practitioners with a structured guide for selecting the appropriate methodology based on the estimand of interest, the maintained assumptions, and the available data.

\section{Empirical Examples}\label{sec:examples}

We now illustrate the empirical applicability of our framework using two examples drawn from recent studies in the persuasion literature. These applications show how the RD persuasion rate and its associated bounds, developed in Section~\ref{sec:iden}, can be estimated using the methods introduced in Section~\ref{sec:estimation}. The first example, based on \citet{brehm2025vaccines}, involves a sharp RD design around a U.S.\ Supreme Court ruling and examines backlash in anti-vaccine newspaper discourse. The second, drawn from \citet{barone2015telecracy}, features a fuzzy RD design and studies how changes in media exposure influenced voting behavior in Italy. Together, these examples demonstrate how our identification and inference results provide practical tools for analyzing persuasion effects in RD designs.  

\subsection{A Vaccine Mandate Ruling and Anti-Vaccine Discourse in Newspapers}

As our first empirical example, we revisit \citet{brehm2025vaccines}, who examine the backlash effect of judicial enforcement of vaccine mandates. They focus on the 1905 U.S.\ Supreme Court ruling in \textit{Jacobson v.\ Massachusetts}, which upheld compulsory smallpox vaccination, and they investigate its impact on anti-vaccine discourse in American newspapers. 
As documented by \citet{harvard2008} and \citet{brehm2025vaccines}, legal scholars have long argued that judicial decisions can provoke public backlash, potentially contributing to the rise of the anti-vaccine movement. These effects may persist beyond the short term. Indeed, \citet{colgrove2006state} shows that average vaccination rates across the United States declined in the decades following the \textit{Jacobson} ruling. We aim to quantify this backlash effect using the RD persuasion rate framework.

The data from \citet{brehm2025vaccines}, available on the journal’s website, are drawn from the Chronicling America Newspaper Archive. The authors initially hand-labeled a random 5\% sample of articles to identify whether each contained anti-vaccination discourse or not. They then used this labeled subset to train a machine learning model to predict anti-vaccine content in the remaining articles. For our analysis, we focus exclusively on the hand-labeled portion of the dataset. See Section~5.5 of \citet{brehm2025vaccines} for further details on the hand-collected \textit{Jacobson} evidence.

In our setup, $Y_{it}$ is a binary variable indicating whether newspaper article $i$ on day $t$ contains anti-vaccine discourse, and $D_{it} = \one(t \geq 0)$, where $t = 0$ denotes the date of the \textit{Jacobson v.\ Massachusetts} ruling. \citet{brehm2025vaccines} exploit a regression discontinuity in time, comparing newspaper articles published just before and just after the ruling.
This setting corresponds to a sharp regression discontinuity design. Following \citet{brehm2025vaccines}, we restrict the sample to a window of 300 days before and after the ruling and, for brevity, consider only one definition of anti-vaccine discourse: whether an article contains anti-vaccine content or not. See the top panel of Figure~A.21 in the Online Appendix of \citet{brehm2025vaccines}. 
Given the 300-day bandwidth, the effective number of observations used in estimation was 813 to the left of the cutoff and 592 to the right.

\begin{table}[!htbp]
\centering 
\caption{\label{tb:anti-vaccine}The RD Persuasion Rates under the Sharp Design} 
\begin{tabular}{lcccc}
\hline\hline
                    &       Right &        Left &  RD &  Persuasion Rate \\
                    &  $\hat{\mu}_+$ & $\hat{\mu}_-$ & $\hat{\mu}_+ - \hat{\mu}_-$ & $(\hat{\mu}_+ - \hat{\mu}_-)/(1 - \hat{\mu}_-)$ \\
\hline
Conventional        &      0.2352   &      0.1837   &      0.0515 &      0.0631 \\
                    &    (0.0348)  &    (0.0259)   &    (0.0433) &    (0.0519) \\\\
Robust Bias-Corrected &  0.2956   &      0.1779   &      0.1177 &      0.1432 \\
                    &    (0.0464)  &    (0.0371)   &    (0.0594) &    (0.0684) \\
\hline
\end{tabular}
\medskip

\begin{minipage}{0.85\textwidth}
{\small \emph{Notes}: 
The table presents estimates for the right- and left-side conditional probabilities at the cutoff, their difference (the standard RD treatment effect), and the RD persuasion rate. The standard errors are in parentheses and are clustered at the state level. The top row shows conventional RD estimates, while the bottom row presents robust bias-corrected estimates using the methods of \citet{calonico2014robust}. All analyses use a symmetric bandwidth of 300 days around the ruling date. The dependent variable, which is human-classified, indicates whether an article contains anti-vaccine discourse. Estimates are computed using the Stata \texttt{rdrobust} package \citep{calonico2017rdrobust}.
\par}
\end{minipage}
\end{table} 

The RD persuasion rate in this context is defined as
\[
\Pr\{ Y_{it}(1) = 1 \mid Y_{it}(0) = 0,\ t = 0 \}.
\]
Here, the assumption of monotonicity, i.e., $Y_{it}(1) \geq Y_{it}(0)$, can be interpreted as follows: if anti-vaccine discourse would have appeared without the ruling, it would also appear with the ruling. To clarify the interpretation of the backlash effect from the ruling, we treat the negative impact of judicial enforcement as the dominant directional effect. 
This monotonicity assumption may be strong in the current setting because the \textit{Jacobson v.\ Massachusetts} ruling could plausibly have caused some newspaper articles (though perhaps not the majority) to shift toward pro-vaccine content. In light of this, we estimate 
$\theta_{\mathrm{RD}}$ as defined in \eqref{eq:srdd}, and interpret it as a lower bound on the true RD persuasion rate, in accordance with our theoretical results (see \cref{lem:ate vs pr} and \cref{thm:sharp}, in particular). Our estimation results are presented in \cref{tb:anti-vaccine}.

\Cref{tb:anti-vaccine} reports the RD persuasion rate estimates under both conventional and robust bias-corrected methods. The conventional estimates indicate a modest increase in anti-vaccine discourse following the ruling, with a persuasion rate of at least 6.3 percent. The bias-corrected estimates yield a larger effect: the RD gap rises to 11.8 percentage points, corresponding to a persuasion rate of at least 14.3 percent. While both estimates provide evidence of backlash, the discrepancy underscores the importance of bias correction in small samples and the use of robust inference procedures.

\subsection{Digital TV Channels in Italian Politics}

For our second example, we revisit \citet{barone2015telecracy}, who study the influence of biased media exposure on political outcomes in Italy. The background of the study is that the Piedmont region had idiosyncratic deadlines for switching to digital television around the time of the 2010 regional elections. Since most analog channels were under Berlusconi’s influence, the switch to digital TV marked a substantial change in media exposure: voters in areas that had transitioned to digital were no longer subject to Berlusconi-biased broadcasting. The western area of Piedmont completed the switch before the 2010 elections, while the eastern area had not.

The outcome variable in \citet{barone2015telecracy} is voting for Berlusconi’s party, and the treatment corresponds to reduced exposure to Berlusconi-slanted media. To align with our notation, we define the outcome to equal one if the individual did \emph{not} vote for Berlusconi’s coalition candidates. The parameter of interest is therefore
\begin{equation}\label{eq:individual}
\Pr\{ Y_{ij}(1) = 1 \mid Y_{ij}(0) = 0,\ W_j = 0 \},
\end{equation}
where $(i,j)$ indexes individual $i$ in town $j$, and $W_j$ denotes the distance between town $j$ and the western/eastern Piedmont border, measured from west to east. Thus, $W_j > 0$ indicates that town $j$ lies in the western area and is assigned to the treatment group.

This setting differs from the previous example in several important ways. First, it constitutes a fuzzy RD design: \( W_j < 0 \) does not imply that none of the individuals in town $j$ had switched to digital TV. In fact, \citet[p.~49]{barone2015telecracy} report:
\begin{quote}
\textit{Access to digital TV in western Piedmont was close to 100 percent in March 2010. About 60 percent of eastern households were on analog TV in March 2010, whereas 40 percent were on digital TV.}
\end{quote}
While this description does not specify how the regions are defined relative to the cutoff, we interpret it as implying \( e(0^+) = 1 \) and \( e(0^-) = 0.4 \) in our notation.

Second, the voting data are aggregated at the town level. That is, the unit of observation is the town, and individual-level votes are not available. The dataset from \citet{barone2015telecracy}, available on the journal’s website, contains information on \( (Y_j, W_j) \), where \( Y_j \) denotes the vote share of Berlusconi’s coalition candidates in town \( j \) in 2010, corresponding to \( \mathbb{E}(Y_{ij} \mid \text{town } j) \). From these data, $\mathbb{E}(Y_{ij} \mid W_j) = \mathbb{E}\bigl\{ \mathbb{E}(Y_{ij} \mid \text{town } j) \ \big| \ W_j\bigr\}$ is identified, but the joint distribution of $(Y_{ij}, D_{ij}, W_j)$ is not.  As a result, this case falls under \cref{thm:fuzzy}(ii) and also under \cref{thm:fuzzy-compliers}(ii).\footnote{Such data limitations are not uncommon in the context of voting, unless the researchers rely on their own survey. Here is another example with similar data limitations. \citet{gerber2011persuasive} study the effects of campaign mailings on turnout and vote shares in the 2006 Kansas election. In that setting, an independent advocacy group sent six pieces of mail criticizing the incumbent Republican attorney general to households selected by a specific algorithm, resulting in a fuzzy RD design. Since voting outcomes were not observed at the household level, the analysis relied on aggregate data at higher levels.}
The former provides bounds on the RD persuasion rate in \eqref{eq:individual}, while the latter applies to the persuasion rate for local compliers:
\[
\Pr\{ Y_{ij}(1) = 1 \mid Y_{ij}(0) = 0,\ \text{$i$ is a local complier} \}.
\]

The main purpose of this example is to illustrate the relevance of our bounds in a fuzzy RD setting and to compare them with the results reported in the existing literature. To that end, rather than conducting new estimation using the original data, we recover inputs from \citet[Section VII.B]{barone2015telecracy} and translate their estimates into our framework. Specifically, they compute a persuasion rate in the style of  \citet{dellavigna2007fox} and \citet{dellavigna2010persuasion}, which they label a ``dissuasion rate'' since the dependent variable is an indicator for voting \emph{for} Berlusconi’s coalition. We reinterpret this as the persuasion rate for \emph{not voting} for Berlusconi’s coalition.
Based on the reported statistics, we infer the following:
\[
\Pr(Y = 1 \mid W = 0^+) = 0.516, \quad
\Pr(Y = 1 \mid W = 0^-) = 0.46, \quad
e(0^+) = 1, \quad
e(0^-) = 0.4.
\]
We treat these sample values as population quantities for the purpose of illustration.\footnote{We explained above how we interpret \( e(0^+) = 1 \) and \( e(0^-) = 0.4 \). The value \( \Pr(Y = 1 \mid W = 0^-) = 0.46 \) is inferred from the statement that “the share of voters in the control area who chose Berlusconi’s coalition in 2010 was 0.54.” The value \( \Pr(Y = 1 \mid W = 0^+) = 0.516 \) is drawn from the claim that “the estimated coefficient for towns 50 km around the border is \(-5.6\) percentage points.”}

By \cref{thm:fuzzy}~(ii), the sharp bounds on \( \Pr\{ Y_{ij}(1) = 1 \mid Y_{ij}(0) = 0,\ W_j = 0 \} \) are:
\begin{align*}
\theta_{\mathrm{RD}} 
&= \frac{0.516 - 0.46}{1 - 0.46}
= \frac{0.056}{0.54}
\approx 0.1037, \\
\theta_{\mathrm{RD},U,e} 
&= \frac{ \min\{1,\ 0.516 + 1 - 1\} - \max\{0,\ 0.46 - 0.4\} }{1 - \max\{0,\ 0.46 - 0.4\}} \\
&= \frac{0.516 - 0.06}{0.94}
= \frac{0.456}{0.94}
\approx 0.4851.
\end{align*}
In addition, by \cref{thm:fuzzy-compliers}~(ii), the lower bound on the persuasion rate for compliers is given by
\begin{align*}
\theta_{cL}^{**} 
&= \max \left\{ \theta_{\mathrm{RD}},\ \frac{0.516 - 0.46}{1 - 0.4} \right\} \\
&= \max\left\{0.1037,\ 0.0933\right\}
= 0.1037,
\end{align*}
with an upper bound of 1. Interestingly, the bounds are tighter without conditioning on compliance behavior.

The dissuasion rate reported in \citet[Section VII.B]{barone2015telecracy} is $-0.203$,\footnote{Specifically, the number $-0.203$ is obtained by 
$\frac{-0.056}{1 - 0.4} \cdot \frac{1}{1 - 0.54}$.} which is based on their definition of the dependent variable as an indicator for voting for Berlusconi's coalition candidates, and the proposal of \citet{dellavigna2007fox}  and \citet{dellavigna2010persuasion}. Adjusting this to reflect the fact that we are treating ``not voting for Berlusconi's candidates'' as the outcome of interest leads to 
\begin{align}
&\frac{ \Pr(Y = 1 \mid W = 0^+) - \Pr(Y = 1 \mid W = 0^-) }{ e(0^+) - e(0^-) }
\cdot \frac{1}{1 - \Pr(Y = 1 \mid W = 0^-)} 
\label{eq:DK}
\\
&= \frac{0.056}{0.6} \cdot \frac{1}{0.54}
\approx 0.1728. 
\notag
\end{align}
However, as in the case of instrumental variable estimation discussed in \citet{jun2023identifying}, it can be shown that the first factor in \eqref{eq:DK} pertains only to local compliers, whereas the second normalizing factor does not. Consequently, the value $0.1728$ cannot be interpreted as a valid persuasion rate for any meaningful subpopulation.

\section{Conclusions}\label{sec:concl}

This paper develops a unified framework for analyzing persuasion effects in RD designs. We introduce the RD persuasion rate, a natural extension of the persuasion rate to settings where treatment assignment depends on whether a running variable crosses a threshold. We establish identification results for both sharp and fuzzy RD designs, accounting for the possibility of limited data availability in the fuzzy case. Estimation and inference can be done by adopting standard RD methods such as local polynomial regression.  We demonstrate the practical value of our results through two empirical applications. Several directions remain for future research. One is to extend the framework to settings with multiple running variables or multi-valued treatments. Another is to integrate RD-based persuasion analysis with machine learning methods to flexibly capture treatment effect heterogeneity across subpopulations.

\begin{appendix}
\section{Proofs}\label{sec:proofs}

\subsection{Lemmas}

We begin by proving several lemmas that will be used in the proofs of the main theorems.

\begin{lemma}\label{lem:ab-full}
Suppose that \cref{ass:mtr,ass:overlap,ass:continuity,ass:threshold} hold, and that $(Y, D, W)$ is jointly observed. Then:
\begin{equation*} 
\Pr(Y = 1 \mid W = 0^+) 
\leq 
\Pr\{ Y(1) = 1 \mid W = 0 \} 
\leq 
\Pr(Y = 1, D = 1 \mid W = 0^+) + 1 - e(0^+).
\end{equation*}
Similarly,
\begin{equation*}
\Pr(Y = 1, D = 0 \mid W = 0^-)
\leq 
\Pr\{ Y(0) = 1 \mid W = 0 \} 
\leq 
\Pr(Y = 1 \mid W = 0^-).
\end{equation*}
All the bounds are sharp.
\proof
Fix  $h > 0$. First, note that $D = \one\{ V \leq e_p(h) \}$ when $W = h > 0$.
Using this fact to write
\begin{equation} \label{eq:Y1}
\Pr\{ Y(1) = 1 \mid W = h \}
= 
\Pr(Y = 1, D = 1 \mid W = h)
+ 
\Pr\{ Y(1) = 1, V > e_p(h) \mid W = h \}.
\end{equation}
We also have that 
\begin{multline}\label{eq:sharp1}
\Pr(Y = 1, D = 0 \mid W = h) 
= 
\Pr\{ Y(0) = 1, V > e_p(h) \mid W = h \} \\
\leq 
\Pr\{ Y(1) = 1, V > e_p(h) \mid W = h \} 
\leq 
1 - e_p(h),
\end{multline}
where the first inequality is due to \cref{ass:mtr}. 

To verify sharpness, let $a_{ij} := \Pr\{ Y(0) = i, Y(1) = j \mid V > e_p(h), W = h \}$. Then, \cref{ass:mtr} implies $a_{10} = 0$, and $a_{00} + a_{01} + a_{11} = 1$. Treating $(a_{01}, a_{11}) \in [0,1]^2$ as free parameters subject to this constraint, we can express the inequalities in \eqref{eq:sharp1} as:
\[
(1 - e_p(h)) a_{11} \leq (1 - e_p(h))(a_{11} + a_{01}) \leq 1 - e_p(h),
\]
which are tight.
Combining \eqref{eq:Y1} and \eqref{eq:sharp1}, we obtain:
\[
\Pr(Y = 1 \mid W = h)
\leq 
\Pr\{ Y(1) = 1 \mid W = h \} 
\leq 
\Pr(Y = 1, D = 1 \mid W = h) + 1 - e_p(h).
\]
Taking the limit as $h \downarrow 0$ and noting that $e_p(0) = e(0^+)$ by continuity yields the first set of inequalities. The second set follows analogously by considering $h < 0$ and is omitted.
\qed
\end{lemma}

\begin{lemma}\label{lem:ab-ecological}
Suppose that \cref{ass:mtr,ass:overlap,ass:continuity,ass:threshold} hold. Further, suppose that $(Y, W)$ is jointly observed and that the function $e$ is known from an external source. Then,
\begin{multline*}
\Pr(Y = 1 \mid W = 0^+)
\leq 
\Pr\{ Y(1) = 1 \mid W = 0 \} \\
\leq 
\min\left\{ 1,\ \Pr(Y = 1 \mid W = 0^+) + 1 - e(0^+) \right\}.
\end{multline*}
Similarly,
\begin{equation*}
\max\left\{ 0,\ \Pr(Y = 1 \mid W = 0^-) - e(0^-) \right\}
\leq 
\Pr\{ Y(0) = 1 \mid W = 0 \}
\leq 
\Pr(Y = 1 \mid W = 0^-).
\end{equation*}
All the bounds are sharp.
\proof
The result follows from \cref{lem:ab-full} combined with the Fr\'{e}chet--Hoeffding inequalities. Specifically, the sharp upper bound on $\Pr\{ Y(1) = 1 \mid W = 0 \}$ is derived using the upper bound in \cref{lem:ab-full} together with the fact that 
\[
\Pr(Y = 1, D = 1 \mid W = 0^+) 
\leq 
\min\left\{ \Pr(Y = 1 \mid W = 0^+),\ \Pr(D = 1 \mid W = 0^+) \right\},
\]
and the sharp lower bound on $\Pr\{ Y(0) = 1 \mid W = 0 \}$ follows similarly from
\[
\max\left\{ 0,\ \Pr(Y = 1 \mid W = 0^-) - e(0^-) \right\}
\leq 
\Pr(Y = 1, D = 0 \mid W = 0^-).   \qedhere
\]
\end{lemma}

\begin{lemma}\label{lem:redundant}
Suppose that \cref{ass:mtr,ass:threshold} hold. Then, for all sufficiently small $h>0$, we have $\Pr(Y=1\mid W=h) \geq \Pr(Y=1\mid W=-h)$.
\proof 
Let $h > 0$ be sufficiently small so that $e_p(h) > e_n(-h)$. Then,
\begin{align*} 
    &\Pr(Y=1\mid W=h) - \Pr(Y=1\mid W=-h) \\
    &=
    \Pr\{ Y(1) = 1, e_n(-h) < V\leq e_p(h) \} - \Pr\{ Y(0) = 1, e_n(-h)< V \leq e_p(h) \} \geq 0.  \qedhere
\end{align*}
\end{lemma}

\subsection{Proofs of the Results in the Main Text}

We now prove the results given in the main text.

\subsection*{Proof of \Cref{lem:ate vs pr}} 

The unconditional version of this result appears as Lemma~1 in \citet{jun2023identifying}. The conditional version stated here follows directly by conditioning on the event $W = 0$, and the proof is therefore omitted. \qed

\subsection*{Proof of \Cref{thm:sharp}}
 
Fix any $h > 0$. Under the sharp RD design, we have $D = \one(W \geq 0)$ almost surely, so $\{W = h\}$ implies $D = 1$ almost surely. Therefore,
\begin{align*}
\Pr(Y = 1 \mid W = h) 
&= \Pr(Y = 1 \mid D = 1, W = h) \\
&= \Pr\{Y(1) = 1 \mid D = 1, W = h\} \\
&= \Pr\{Y(1) = 1 \mid W = h\}.
\end{align*}
Taking the limit as $h \downarrow 0$ and applying \cref{ass:continuity}, we obtain
\[
\Pr(Y = 1 \mid W = 0^+) = \Pr\{Y(1) = 1 \mid W = 0\}.
\]
A symmetric argument applies for $h < 0$, where $D = 0$ almost surely. Thus,
\[
\Pr(Y = 1 \mid W = 0^-) = \Pr\{Y(0) = 1 \mid W = 0\}.
\]
Substituting these expressions into the definition of $\theta_{\mathrm{RD}}$ in \eqref{eq:srdd}, we have $\theta_L = \theta_{\mathrm{RD}}$. The final claim follows immediately from \cref{lem:ate vs pr}.
\qed

\subsection*{Proof of \Cref{thm:fuzzy}} We focus on the upper bound: the lower bound is similar but simpler, because it depends only on the distribution of $(Y,W)$.  For the first assertion, we consider 
\[
\max_{a,b}\  \frac{a-b}{1-b}  \quad \text{subject to}\quad m_a \leq a\leq M_a,\ m_b\leq b\leq M_b,\ a\geq b,
\]
where $m_a,m_b, M_a$, and $M_b$ are given in \cref{lem:ab-full}. Here, the constraint $a\geq b$ is redundant because $m_a\geq M_b$ by \cref{lem:redundant}. Therefore, concentrating out $a$, and then using the fact that $(M_a - b)/(1-b)$ is decreasing in $b$ yields the maximum value equal to $(M_a - m_b)/(1-m_b)$. The sharpness is a consequence of continuity and the intermediate value theorem.   The second assertion follows by the same reasoning but by using \cref{lem:ab-ecological} instead of \cref{lem:ab-full}. The third assertion follows by maximizing $\theta_{RDD,U_e}$ with respect to $0\leq e(0^-)\leq e(0^+) \leq 1$. \qed 

\subsection*{Proof of \Cref{thm:fuzzy-compliers}}

We begin with the point identification result in part (i). By \cref{ass:threshold}(ii), the conditional expectation of $D$ given $W$ can be written as
\[
\Exp[D \mid W = \pm h] = \Exp[D(\pm h)].
\]
Using \cref{ass:threshold}(i), the difference in these expectations becomes
\[
    \Exp[D(h)] - \Exp[D(-h)] = \mathbb{P}[D(h) - D(-h) = 1],
\]
which equals the proportion of compliers at the threshold.

Next, decompose the observed outcome: for sufficiently small $h>0$,
\begin{align*}
\Exp[Y \mid W = h] 
&= \Exp[ Y(0) \mid W = h] + \Exp[ D\{Y(1) - Y(0) \} \mid W = h] \\
&= \Exp[Y(0)] + \Exp[D(h)\{Y(1) - Y(0) \}],
\end{align*}
where the final line uses \cref{ass:threshold}(ii). Similarly,
\[
    \Exp[Y \mid W = -h] = \Exp[Y(0)] + \Exp[D(-h)\{Y(1) - Y(0)\}].
\]
Subtracting, we obtain:
\[
    \Exp[Y \mid W = h] - \Exp[Y \mid W = -h]
    = 
    \Exp[ \{ D(h) - D(-h) \}\{ Y(1) - Y(0) \} ].
\]
It follows that
\[
    \Exp[ Y(1) - Y(0) \mid D(h) - D(-h) = 1] 
    = 
    \frac{\Exp[Y \mid W = h] - \Exp[Y \mid W = -h]}
        {\Exp[D \mid W = h] - \Exp[D \mid W = -h]}.
\]
Taking the limit as $h \downarrow 0$, we obtain:
\[
\Pr\{ Y(1) = 1 \mid \sC_0 \} - \Pr\{ Y(0) = 1 \mid \sC_0 \} 
= 
\frac{\Pr(Y = 1 \mid W = 0^+) - \Pr(Y = 1 \mid W = 0^-)}{e(0^+) - e(0^-)}.
\]
Hence,
\begin{align}
\theta_{cL}
&= 
\frac{\Pr(Y = 1 \mid W = 0^+) - \Pr(Y = 1 \mid W = 0^-)}{(e(0^+) - e(0^-)) \Pr\{ Y(0) = 0 \mid \sC_0 \}} \notag \\
&= 
\frac{\Pr(Y = 1 \mid W = 0^+) - \Pr(Y = 1 \mid W = 0^-)}{\Pr\{ Y(0) = 0,\ e(0^-) < V \leq e(0^+) \}}.
\label{eq:thetacL0}
\end{align}

To identify the denominator, observe:
\begin{align}
\Exp[ (1 - Y)(1 - D) \mid W = h ] 
&= 
1 - \Exp[Y(0)] - \Exp[ D(h)\{1 - Y(0)\} ], \notag 
\\
\Exp[ (1 - Y)(1 - D) \mid W = -h] 
&= 
1 - \Exp[Y(0)] - \Exp[ D(-h)\{1 - Y(0)\} ].
\label{eq:y0d0}
\end{align}
Subtracting yields:
\begin{align}
&\mathbb{E}[(1 - Y)(1 - D) \mid W = h] - \mathbb{E}[(1 - Y)(1 - D) \mid W = -h] \notag \\
&= -\mathbb{E}[(D(h) - D(-h))(1 - Y(0))].
\label{eq:den1}
\end{align}
Taking the limit as $h \downarrow 0$ and combining \cref{eq:thetacL0,eq:den1} completes the proof of part (i).

For part (ii), consider again \cref{eq:thetacL0}. We seek sharp bounds on $\Pr(Y_0 = 0,\ \sC_0)$ in the case where $(Y, D)$ is not jointly observed. First, note:
\begin{align*}
\Pr(Y = 1 \mid W = 0^+) - \Pr(Y = 1 \mid W = 0^-) 
&= \Pr\{ Y(0) = 0,\ Y(1) = 1,\ \sC_0 \} \\
&\leq \Pr\{ Y(0) = 0,\ \sC_0 \},
\end{align*}
where the inequality is sharp under the Fréchet–Hoeffding bound and \cref{ass:mtr}. Thus, the upper bound on $\theta_{cL}$ is 1.

To sharpen the lower bound, we bound $\Pr\{ Y(0) = 0,\ \sC_0 \}$ above. First,
\begin{align*}
\Pr\{ Y(0) = 1,\ V > e(0^-) \} 
&= \lim_{h \downarrow 0} \Exp[ \{1 - D(-h)\}\{ 1 - Y(0) \} ] \\
&= \lim_{h \downarrow 0} \Pr(Y = 0,\ D = 0 \mid W = h) \\
&\leq \min\bigl\{ \Pr(Y = 0 \mid W = 0^-),\ 1 - e(0^-) \bigr\},
\end{align*}
using \cref{eq:y0d0} and again applying the Fréchet–Hoeffding inequality. Also, since $\Pr\{ Y(0) = 1,\ V \leq e(0^+) \} \leq e(0^+)$, we obtain:
\[
\Pr\{ Y(0) = 0,\ \sC_0 \} 
\leq 
\min\bigl\{ \Pr(Y = 0 \mid W = 0^-),\ e(0^+) - e(0^-) \bigr\}.
\]
This yields the sharp lower bound:
\[
\theta_{cL}^{**} 
:= 
\max\left\{ \theta_{\mathrm{RD}},\ \frac{\Pr(Y = 1 \mid W = 0^+) - \Pr(Y = 1 \mid W = 0^-)}{e(0^+) - e(0^-)} \right\}.
\]

Finally, for part (iii), if no information about $e$ is available, then $e(0^+) - e(0^-)$ could equal 1. In that case, the bound reduces to $[\theta_{\mathrm{RD}},\ 1]$. \qed

\subsection*{Remark on Sharpness in \Cref{thm:fuzzy-compliers}}

The sharpness result in the second part of \cref{thm:fuzzy-compliers} can be verified more directly as follows. Consider \cref{tb:jprob} in which the joint probabilities of the potential outcomes $(Y(0), Y(1))$ and the latent variable $V$ are grouped into three intervals based on the value of $V$.

\begin{table}[!htbp]
\centering
\begin{tabular}{c|cc|c}
Group & $Y(0)$ & $Y(1)$ & Joint Probability \\
\hline
$V \leq e^-$ & 0 & 0 & $a_1$ \\
$V \leq e^-$ & 0 & 1 & $b_1$ \\
$V \leq e^-$ & 1 & 1 & $c_1$ \\
\hline
$e^- < V \leq e^+$ & 0 & 0 & $a_2$ \\
$e^- < V \leq e^+$ & 0 & 1 & $b_2$ \\
$e^- < V \leq e^+$ & 1 & 1 & $c_2$ \\
\hline
$e^+ < V \leq 1$ & 0 & 0 & $a_3$ \\
$e^+ < V \leq 1$ & 0 & 1 & $b_3$ \\
$e^+ < V \leq 1$ & 1 & 1 & $1 - \sum_{i=1}^3 (a_i + b_i) -  \sum_{i=1}^2 c_i$ \\
\end{tabular}
\caption{Joint Probabilities of Potential Outcomes and Latent Type}
\label{tb:jprob}
\end{table}

From the observed data, we obtain the following quantities:
\begin{align*}
&e^+ := \Pr(D=1 \mid W=0^+), \quad
e^- := \Pr(D=1 \mid W=0^-), \\
&p^+ := \Pr(Y=1 \mid W=0^+), \quad
p^- := \Pr(Y=1 \mid W=0^-).
\end{align*}
These imply the following constraints on the joint probabilities in \Cref{tb:jprob}:
\begin{align}
a_1 + b_1 + c_1 &= e^-, \label{eq1} \\
a_2 + b_2 + c_2 &= e^+ - e^-, \label{eq2} \\
a_1 + a_2 + a_3 + b_3  &= 1 - p^+, \label{eq3} \\
a_1 + a_2 + a_3 + b_2 + b_3 &= 1 - p^-. \label{eq4}
\end{align}
\Cref{eq1,eq2} follow directly from the definitions of $e^+$ and $e^-$, representing the total probability mass of treated individuals just above and just below the threshold, respectively. \Cref{eq3} corresponds to the observed probability that \( Y = 0 \) given \( W = 0^+ \), that is, \( \Pr(Y = 0 \mid W = 0^+) = 1 - p^+ \). The relevant types include: (i) individuals for whom both \( Y(0) = 0 \) and \( Y(1) = 0 \), across all regions of \( V \) (i.e., \( a_1, a_2, a_3 \)); and (ii) individuals with \( Y(0) = 0 \) and \( Y(1) = 1 \) who are untreated at \( W = 0^+ \) (i.e., \( b_3 \)). These individuals are all observed with \( Y = 0 \) at \( W = 0^+ \), which justifies the constraint. Similarly, \Cref{eq4} reflects \( \Pr(Y = 0 \mid W = 0^-) = 1 - p^- \). The contributing groups are: (i) individuals with \( Y(0) = Y(1) = 0 \) across all regions (i.e., \( a_1, a_2, a_3 \)); and (ii) individuals with \( Y(0) = 0 \), \( Y(1) = 1 \), who are untreated at \( W = 0^- \) (i.e., \( b_2, b_3 \)). These classifications fully account for those observed with \( Y = 0 \) at \( W = 0^- \), validating the constraint.

We seek bounds on the quantity $\Pr\{ Y(0) = 0,\ e^- < V \leq e^+ \} = a_2 + b_2$. Subtracting \eqref{eq4} from \eqref{eq3} gives:
\[
b_2 = p^+ - p^-.
\]
Substituting this into \eqref{eq2} gives:
\[
a_2 + c_2  = e^+ - e^- - p^+ + p^-.
\]
We now have:
\begin{align*}
b_2 &= p^+ - p^-, \\
b_1 &= e^- - a_1 - c_1 \geq 0, \\
c_2 &= e^+ - e^- - p^+ + p^- - a_2 \geq 0, \\
a_3 &= 1 - p^+ - a_1 - a_2 - b_3 \geq 0.
\end{align*}
This gives rise to the following constraints:
\begin{align}
0 &\leq a_1 + c_1 \leq e^-, \label{eq:info1} \\
0 &\leq a_2 \leq e^+ - e^- - p^+ + p^-, \label{eq:info2} \\
0 &\leq a_1 + a_2 + b_3 \leq 1 - p^+. \label{eq:info3}
\end{align}
Now observe that:
\[
a_2 + b_2 = a_2 + (p^+ - p^-) \quad \Rightarrow \quad a_2 + b_2 \geq p^+ - p^-
\]
since $a_2 \geq 0$.
From \eqref{eq:info2}, we know:
\[
a_2 \leq e^+ - e^- - (p^+ - p^-),
\]
so
\[
a_2 + b_2 \leq e^+ - e^-.
\]
Also, from \eqref{eq4}:
\[
a_2 + b_2 + a_3 + b_3 + a_1 = 1 - p^- \quad \Rightarrow \quad a_2 + b_2 \leq 1 - p^-,
\]
since all other components are nonnegative.
Thus, putting everything together, we have:
\begin{align}\label{eq:sharp:bound:a2b2}
p^+ - p^- \leq a_2 + b_2 \leq \min\{ e^+ - e^-,\ 1 - p^- \},
\end{align}
and this interval coincides with the bounds for \( \Pr\{ Y(0) = 0,\ \sC_0 \} \) derived in \cref{thm:fuzzy-compliers}.

To establish the sharpness of the bounds given in \eqref{eq:sharp:bound:a2b2}, we must show that every value within this interval, including the endpoints, can be attained by some admissible configuration of the joint probabilities in \Cref{tb:jprob} that satisfies the constraints in \Cref{eq1,eq2,eq3,eq4}. Let $\alpha \in [p^+ - p^-,\ \min\{ e^+ - e^-,\ 1 - p^- \}]$ be arbitrary. Fix $b_2 = p^+ - p^-$ and choose $a_2 = \alpha - b_2$, which is non-negative by construction. Next, we set the values of $(a_1, c_1, b_3)$ to satisfy the constraints in \cref{eq:info1,eq:info2,eq:info3}. Thus, for any value in the identified interval, one can construct a compatible joint distribution satisfying all constraints and achieving the given value $a_2 + b_2 = \alpha$, which establishes the sharpness of the bounds. \qed

\end{appendix}  

    
    \bibliographystyle{ecta-fullname} 
    
    \bibliography{persuasion4}  

        
        


        

\end{document}